\documentclass[12pt]{iopart}
\usepackage{bbold}
\def\be{\begin{equation}}
\def\ee{\end{equation}}
\def\ba{\begin{array}}
\def\ea{\end{array}}
\def\e{\kappa}

\def\s{\sigma}

\def\l{\lambda}
\def\ei{\kappa_i}
\def\ej{\kappa_j}
\def\T2{T^{(2)}}

\begin{document}

\comment[Comment on Kuzovkov \etal]{Comment on `Exact analytical
solution for the generalized 
Lyapunov exponent of the two-dimensional Anderson localization'}

\author{P. Marko\v{s}\dag\ddag\, L. Schweitzer\dag\, and M. Weyrauch\dag}

\address{\dag\ Physikalisch-Technische Bundesanstalt, D-38116 Braunschweig, Germany}

\address{\ddag\ Institute of Physics, Slovak Academy of Sciences, 845
11 Bratislava, Slovakia}

\begin{abstract}
In a recent publication, \JPCM {\bf 14} 13777 (2002), 
Kuzovkov \etal announced an
analytical solution of the two-dimensional Anderson localisation
problem via the calculation of a generalised Lyapunov exponent using
signal theory. Surprisingly, for certain energies and small disorder
strength they observed delocalised states. 
We study the transmission properties of the same model using well-known
transfer matrix methods. Our results disagree with the findings obtained using
signal theory. We point to the possible origin of this discrepancy and
comment on the general strategy to use a generalised Lyapunov exponent
for studying Anderson localisation.
  
\end{abstract}

\submitto{\JPCM}


\maketitle

It is generally believed that in the absence of both spin-orbit scattering
and magnetic fields, there is no metal-insulator transition for non-interacting
electrons in disordered two-dimensional (2D) systems. 
This belief is based on the scaling hypothesis \cite{AALR} and
supported by detailed finite-size scaling analysis of numerical data \cite{McKK}.
Nevertheless, an analytical proof that there are no extended states in
a disordered 2D system described by the Anderson model is still missing.   

Recently, Kuzovkov \etal \cite{K} studied disordered 2D systems
using signal theory, and they found parameter regimes where the system under 
consideration transmits the incident signal. They interpret this
result as an indication for a metallic phase. This strongly
contradicts standard wisdom, and, therefore, in this Comment, we
reconsider the transmission properties of a 2D Anderson model using
well-known transfer matrix methods. 
Similar calculations were
presented before by Pendry \cite{P}.  

Consider a 2D Anderson model on the $M \times L$ lattice ($M$ is the width and
$L$ is the length of the system).  The discrete Schr\"odinger equation 
\be\label{sche}
\psi_{n+1,m}=(E-\varepsilon_{nm})\psi_{nm}-\psi_{n-1,m}-\psi_{n,m+1}-\psi_{n,m-1}
\ee
may be rewritten using the transfer matrix $T^{(1)}_n$ as
\be\label{tme}
\left(
\ba{l}
\Psi_{n+1}\\
\Psi_n
\ea
\right)
=T^{(1)}_n
\left(
\ba{l}
\Psi_n\\
\Psi_{n-1}
\ea
\right),~~~~~
T^{(1)}_n=\left(
\begin{array}{rr}
E-H_0-\epsilon_n  &  -\mathbb{1}\\
\mathbb{1}      &   \mathbb{0}
\end{array}
\right).
\ee
Here, $H_n=H_0+\epsilon_n$ is the ($M\times M$) Hamiltonian of the $n$th slice
which contains random (uncorrelated) energies $\varepsilon_{nm}$,
and $\Psi_n$ is the vector $(\psi_{n1},\psi_{n2},\dots,\psi_{nM})$. 
For simplicity we assume for the disorder potentials that
$\langle \varepsilon_{nm} \rangle= 0$ and $\langle 
\varepsilon_{nm}\varepsilon_{n^\prime m^\prime} 
\rangle=\sigma^2\delta_{nn^\prime}\delta_{mm^\prime}$.
The angle brackets denote averages over disorder.

In order to study the properties of quantities $\langle \psi_{nm} \psi^*_{nm}\rangle$ 
(signals) as considered by Kuzovkov \etal \cite{K} we
construct the tensor product of Eq.~(\ref{tme}),
\be\label{tme2}
\left(
\ba{l}
\Psi_{n+1}\\
\Psi_n
\ea
\right)
\otimes
\left(
\ba{l}
\Psi^*_{n+1}\\
\Psi^*_n
\ea
\right)
=\T2_n
\left(
\ba{l}
\Psi_n\\
\Psi_{n-1}
\ea
\right)
\otimes
\left(
\ba{l}
\Psi^*_n\\
\Psi^*_{n-1}
\ea
\right).
\ee
The size of the matrix $\T2_n=T_n^{(1)}\otimes T_n^{(1)}$ is  $4M^2\times 4M^2$. 
We may now average Eq.~(\ref{tme2}) over the disorder. Since the matrix $\T2_n$ 
and the tensor product on which it operates are statistically independent, 
we obtain a non-random matrix
$\T2=\langle \T2_n\rangle$,
\be\label{tm2o}
\T2=
\left(
\ba{rrrr}
\s^2 \mathbb{1}\otimes \mathbb{1}+D_0\otimes D_0 &  -D_0 \otimes
\mathbb{1}  &  -\mathbb{1} \otimes D_0  &  \mathbb{1}\\
D_0\otimes \mathbb{1}       &   \mathbb{0}  &   -\mathbb{1}  &  \mathbb{0}\\
\mathbb{1} \otimes D_0      &   -\mathbb{1} &    \mathbb{0}  &  \mathbb{0}\\
\mathbb{1}                  &    \mathbb{0} &    \mathbb{0}  &  \mathbb{0}
\ea
\right)
\ee
with $D_0=E-H_0$. The matrix (\ref{tm2o}) can be
transformed by the matrix $Q\otimes Q$, where $Q$ diagonalises $H_0$.
Using Dirichlet boundary conditions in the transversal direction, one obtains
\be
Q D_0Q^{-1}=\e,~~~~\e_i=E-2\cos k_i,~~~~~k_i=\frac{\pi}{M+1}i,~~~i=1,\ldots,M.
\ee
As a result, the $4M^2$ eigenvalues $\lambda$ of 
$\T2$ may be calculated from
\be\label{xxy}
\prod_{i,j}^M\det\left(
\ba{rrrr}
\ei\ej+\s^2-\lambda  &  -\ei  &  -\ej  &  1\\
\ei       &  -\lambda         &   -1   &  0\\
\ej      &   -1         &    -\lambda   &  0\\
1       &      0         &   0     &  -\lambda
\ea
\right)=0.
\ee
Therefore, the eigenvalues fulfil the equations
\be\label{equation}
\l^4-\l^3(\ei\ej+\s^2)+\l^2(\ei^2+\ej^2-2)-\l(\ei\ej-\s^2)+1=0.
\ee
Using $\l=\exp(iq)$, the Eqs.~(\ref{equation}) read
\be\label{eq}
2\cos 2q -2\ei\ej\cos q +(\ei^2+\ej^2-2) =2\s^2i\sin q.
\ee
For the ordered case ($\sigma=0$) the $4M^2$ solutions are obtained as
$q_{ij}=\pm a_i\pm a_j$ with $2\cos a_i=E-2\cos k_i$, and  
the eigenvalues of the unperturbed matrix are given by
\be\label{zero}
\l(\s^2=0)=\exp\left[i(\pm a_i\pm a_j)\right].
\ee
If $\sigma \neq 0$, then Eq.~(\ref{eq}) shows that $q$  is real only
(i.e. $|\lambda|=1$),  if  
$q$ is zero or a multiple of $\pi$. Therefore, the only
eigenvalues  of the transfer matrix $\T2$ which lie on the unit circle
are $\l=\pm 1$.
Note that these eigenvalues are independent of the
disorder. Therefore, it is evident that they correspond to some
internal symmetry of the Anderson model (see also Ref. \cite{P}). 
In fact, from Eqs.~(\ref{equation}) one finds $M$ eigenvalues
$\lambda=1$ (one for each  $i=j$). 
They just correspond to current conservation in each of the $M$
channels \cite{P}. We will call these eigenvalues `trivial
eigenvalues'. 
Eigenvalues $\l=-1$ are obtained for $\ei=-\ej$ corresponding to the special energy
$E=\cos k_i +\cos k_j$, which can be discarded as irrelevant.

Our results so far do not disagree with those of
Ref.~\cite{K}. However, Kuzovkov \etal claim 
that in the limit $M \rightarrow \infty$ eigenvalues $|\lambda|=1$ will
appear, which do not correspond to the trivial eigenvalues discussed above.  
Mathematically, the limit $M\rightarrow\infty$ of the discrete model
discussed here bears various conceptual and technical
difficulties. Therefore we studied this limit numerically and find no
new solutions with $|\lambda|=1$ apart from the trivial ones. 
It therefore appears that the metallic solutions found by Kuzovkov
\etal are an artefact of a questionable mathematical limit procedure. 

In conclusion, our analysis shows that apart from the `trivial' 
eigenvalues 
$\l=\pm 1$, 
which correspond to symmetries of the 2D Anderson Hamiltonian,
all other eigenvalues have an absolute value different from
unity. This holds for any $M$. 
Therefore, all input signals 
are damped away if the system is sufficiently long.  
This result contradicts the findings in Ref. \cite{K}. 
We believe that the reasons  
for this contradiction are as follows:  
Kuzovkov \etal reduced the size of the transfer matrix from
$4M^2\times 4M^2$ to only $\sim M$. 
This drastic reduction was achieved by introducing 
a translational symmetry in the transversal direction. 
A precise description and motivation of this additional averaging procedure is absent 
in their paper. We do not believe that such a symmetry really holds in general. 
We therefore have to conclude that by this reduction of the size of
the transfer matrix they transformed the model defined in
Eq.~(\ref{sche}) into another model. 
But even then Kuzovkov \etal do not find eigenvalues $|\lambda|=1$ different from the
trivial eigenvalues. Only using a questionable limit procedure $M \rightarrow \infty$
such eigenvalues are obtained.

Finally, we would like to comment on the general strategy to tackle the
Anderson localization problem using methods similar to the one
discussed here and in Ref.~\cite{K}.
We do not think that the analysis of the second moments
of the wave function $|\psi|^2$ (or of any higher moments
$|\psi|^{2m}$) as was done here as well as in Ref.~\cite{K} 
is able to detect a metallic phase.  The physical reason is the
following:
In order to detect a metallic phase one needs to average the logarithm
of the wave function itself \cite{PS81}. The procedures discussed here, while
mathematically correct, always 
lead to quantities which are damped away for large systems, i.e. 
yield eigenvalues $|\lambda|\neq 1$ apart from the trivial ones.
This can be exemplified most drastically with an analysis of a 3D
system: Here, one obtains Eq.~(\ref{xxy}) with $\e_i=E-2\cos k_\alpha -2\cos
k_\beta$, $i=M(\alpha-1)+\beta$, $\alpha,\beta=1,\dots,M$. Consequently,
the absolute values of all non-trivial eigenvalues differ from unity
independently of the strength of the 
disorder. Since weakly disordered 3D systems are metallic, we conclude
that in contrast to $\langle\ln(|\psi|)\rangle$ an analysis of $\langle|\psi|^2\rangle$
does not help to observe a metallic regime even if it exists.

\bigskip

\noindent{\bf Acknowledgement} PM thanks PTB for financial support
which enabled his stay at PTB; furthermore he acknowledges support
from grant VEGA 2/3108/2003.

\end{document}